\documentclass[aps,amsmath,amssymb,twocolumn,floatfix]{revtex4-2}

\usepackage{graphicx}% Include figure files
\usepackage{dcolumn}% Align table columns on decimal point
\usepackage{bm}% bold math
\begin{document}

\title{Antiferromagnetism of CeCd$_{0.67}$As$_{2}$ existing deep inside the narrow gap semiconducting state}% Force line breaks with \\

%\thanks{A footnote to the article title}%

\author{Suyoung Kim$^1$, Obinna P. Uzoh$^1$, and Eundeok Mun$^{1,2}$}

%\email{emun@sfu.ca}
\affiliation{$^1$ Department of Physics, Simon Fraser University, Burnaby, BC V5A 1S6 Canada}
\affiliation{$^2$ Center for Quantum Materials and Superconductivity (CQMS), Sungkyunkwan University, Suwon 16419, South Korea}

%\date{\today}

\begin{abstract}

Single crystals of $R$Cd$_{0.67}$As$_2$ ($R$ = La and Ce) have been synthesized by high temperature ternary melt and their physical properties have been explored by means of magnetization, specific heat, electrical resistivity, Hall coefficient, and thermoelectric power measurements. $R$Cd$_{0.67}$As$_2$ compounds indicate a (structural) phase transition at high temperatures, accompanied by a remarkable increase of the electrical resistivity with an extremely low carrier concentration. CeCd$_{0.67}$As$_2$ exhibits a large magnetic anisotropy and an antiferromagnetic (AFM) order below $T_{N} = 4$~K. Magnetic susceptibility curves, together with magnetization isotherms and specific heat, are analyzed by the point charge model of crystalline electric field (CEF). In the paramagnetic state, the observed magnetic properties can be well explained by the CEF effects, implying that the 4$f$ moments remain localized. Electrical resistivity measurements, together with Hall resistivity and thermoelectric power, also suggest highly localized 4$f$ electrons, where Kondo contributions are negligible. The low temperature physical properties manifest strong magnetic field dependencies. For $H \perp c$, $T_{N}$ shifts to lower temperature as magnetic field increases, and eventually disappears at $H_{c} \sim 60$~ kOe. Inside the AFM state, three metamagnetic transitions are clearly evidenced from the magnetization isotherms. The RKKY interaction may be responsible for the AFM ordering in CeCd$_{0.67}$As$_2$, however it would have to be mediated by extremely low charge carriers. Although the AFM ordering temperature in CeCd$_{0.67}$As$_2$ can be continuously suppressed to zero, no AFM quantum phase transition is expected due to the lack of conduction electron clouds to screen the 4$f$ moments.

\end{abstract}

%\keywords{Suggested keywords}%Use showkeys class option if keyword
                              %display desired
\maketitle

%\tableofcontents

\section{Introduction}

Ce-based intermetallic compounds have shown a variety of interesting phenomena such as heavy fermion (HF) behavior~\cite{Stewart1984}, valence fluctuation~\cite{Lawrence1981}, unconventional superconductivity~\cite{Steglich1979, Ounki2004, Settai2007, Pfleiderer2009}, and quantum phase transitions~\cite{Gegenwart2008, Coleman2010, Weng2016}. These interesting phenomena are related to the hybridization between localized 4$f$ electrons and conduction electrons. A strong hybridization delocalizes the 4$f$ electron pushing the system into a paramagnetic Kondo state, whereas a weaker hybridization results in an antiferromagnetic (or ferromagnetic) order via Ruderman-Kittel-Kasuya-Yosida (RKKY) interaction.
When the Doniach-like diagram is considered, a general assumption is that the system contains sufficiently large amount of conduction electrons to screen the local moments and to mediate a coupling between the localized 4$f$ moments. An interesting question is how does the low carrier density affect these phenomena ~\cite{Nozieres1998, Pruschke2000}. Because the low carrier density means a lack of conduction electrons to either screen the moments or mediate the interaction between moments, both RKKY and Kondo interactions are expected to be weakened. There are only a small fraction of low carrier density 4$f$ HF systems such as CeNiSn, CeRhSn, Ce$_{3}$Pt$_{3}$Bi$_{4}$, CeNi$_{2-x}$As$_{2}$, and Yb$_{3}$Ir$_{4}$Ge$_{13}$~\cite{Takabatake1996, Nishigori1996, Hundley1990, Luo2015, Chen2019, Rai2019}. Strongly correlated electronic states in low-carrier Kondo lattice systems, more recently exploring also unconventional quantum criticality, continue to attract considerable attention ~\cite{Riseborough2000, Tokiwa2015, Luo2015, Chen2019, Dzsaber2017, Rai2019}. To enhance our knowledge in such a regime studies of more materials are needed.

We have systematically studied the physical properties of $R$Cd$_{0.67}$As$_{2}$ ($R$ = La and Ce) system to investigate RKKY and Kondo interactions deep inside the low carrier, semiconducting state. $R$Cd$_{0.67}$As$_{2}$ compounds belong to a large number of ternary pnictides $RTX_{2}$ ($R$ = rare-earth, $T$ = transition metal, $X$ = As, Sb, and Bi)~\cite{Brylak1995, LeitheJasper1994, Sologub1995, Myers1999, Stoyko2011, Petrovic2003}. Many of these compounds typically adopt the tetragonal HfCuSi$_{2}$-type structure ($P4/nmm$, Z = 2, no 129). Among these series the ternary antimonides with $T$ = Ag showed that all atomic positions are fully occupied \cite{Brylak1995}, while a deficiency of the transition metal site frequently occurs, corresponding to the formula $R$$T_{1-x}$$X_{2}$, due to the distortion of $X$ square nets~\cite{Cordier1985, Nientiedt1999, Eschen2003, Stoyko2011}. In addition, an excess of the transition metal site can also be observed, as seen in $R$Cu$_{1+x}$As$_{2}$~\cite{Wang1999}. In the ternary arsenides $RT$As$_{2}$, crystal structure and physical properties have been previously investigated for $T$ = Cu~\cite{Brylak1995}, Cd~\cite{Piva2021}, and Zn~\cite{Stoyko2011}. The $R$CuAs$_{2}$ compounds form a tetragonal HfCuAs$_{2}$-type structure and show an anomalous transport property~\cite{Sampathkumaran2003, Evans2022}. It has been reported that LaZn$_{0.67}$As$_{2}$ crystallizes into a body centered tetragonal SrZnBi$_{2}$-type structure ($I4/mmm$, Z = 4, no 139), while $R$Zn$_{1-x}$As$_{2}$ ($R$ = Ce$-$Sm) form a defect HfCuSi$_{2}$-type structure \cite{Stoyko2011}. In particular, the structural refinement of PrZn$_{0.67}$As$_{2}$ found an ordering of a vacant Zn site, forming an ordered superstructure of Pr$_{3}$Zn$_{2}$As$_{6}$ ($Pmmn$, Z = 2, no 59)~\cite{Nientiedt1999}. Interestingly, the electrical resistivity of LaZn$_{0.67}$As$_{2}$ shows an activated behavior (semiconductor-like behavior) arising from an energy gap \cite{Stoyko2011}.

In this manuscript, we report the crystal structure, thermodynamic, and transport properties of $R$Cd$_{0.67}$As$_{2}$ ($R$ = La and Ce). Single crystals of these compounds have been synthesized by high temperature ternary melt with excess of Cd and As. At high temperatures, physical property measurements confirm a (structural) phase transition ($T_{s}$ = 280~K for $R$ = La and $T_{s}$ = 137~K for $R$ = Ce) and transport property measurements point to a semiconducting state below $T_{s}$ with a low carrier density. The observed high temperature magnetic property of CeCd$_{0.67}$As$_{2}$ can be well understood in the framework of crystalline electric field (CEF) effect. We demonstrate that the CEF ground state of Ce 4$f$ electron is an antiferromagnetic doublet with $T_{N}$ = 4~K. It is of interest to explore antiferromagnetism existing deep inside the narrow gap semiconducting phase to see whether they are similar to or different from those found in good metallic systems. Recently, the existence of nonstoichiometric $R$Cd$_{0.67}$As$_{2}$ ($R$ = La and Ce) and their physical properties have been reported~\cite{Piva2021, Piva2022}, where the compounds are synthesized in single crystalline form by chemical vapor transport (CVT) technique. Both La- and Ce-based compounds crystallize into a vacancy ordered superstructure of the body centered tetragonal ($I4/mmm$) parent structure, corresponding to $R_{3}$Cd$_{2}$As$_{6}$ stoichiometry. As temperature is decreased, both compounds display the phase transition at $T$ = 136~K and 278~K for Ce$_{3}$Cd$_{2}$As$_{6}$ and La$_{3}$Cd$_{2}$As$_{6}$, respectively, where these phase transitions are attributed to the charge density wave (CDW) driven structural distortions. Ce$_{3}$Cd$_{2}$As$_{6}$ exhibits an antiferromagnetic ordering below $T_{N}$ = 4~K which increase to 5.3~K under external pressure of 3.8~GPa \cite{Piva2022}.

\section{Experiments}

\begin {figure}
\includegraphics[width=1\linewidth]{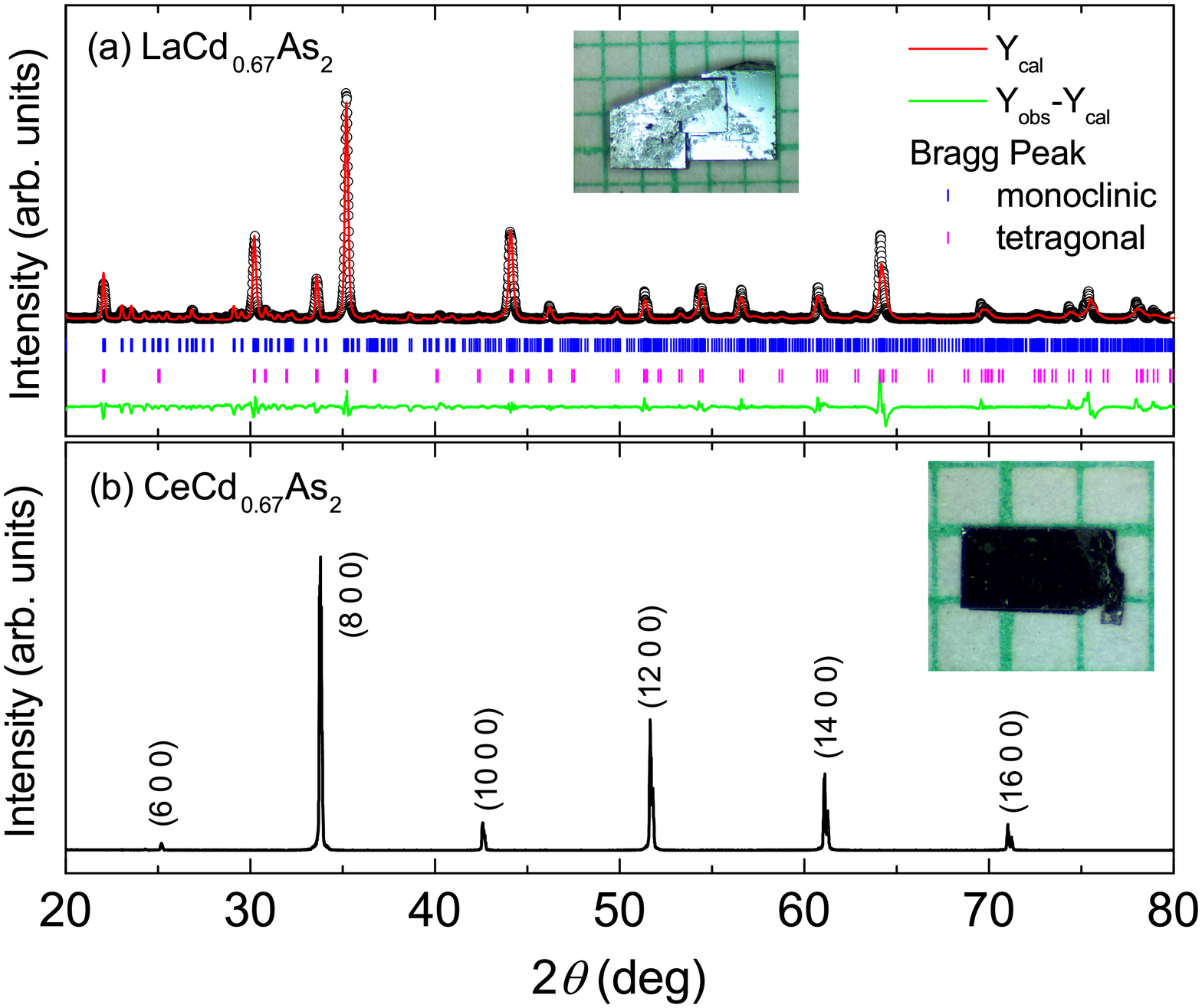}
\caption {(a) Powder XRD patterns of LaCd$_{0.67}$As$_{2}$. Red line is the calculated XRD patterns with a monoclinic structure ($C2/m$). Green lines represents the difference between the data and calculated XRD patterns. The Bragg positions are shown for both monoclinic (top) and tetragonal (bottom) structure. Inset shows a photograph of as grown LaCd$_{0.67}$As$_{2}$ single crystal on mm grid. (b) Single crystal X-ray patterns for CeCd$_{0.67}$As$_{2}$. The peak indices ($h$, 0, 0) are in the monoclinic structure. Inset show a photo of CeCd$_{0.67}$As$_{2}$ over the mm scale.}
\label{FIG1}
\end{figure}

Single crystals of $R$Cd$_{0.67}$As$_{2}$ ($R$ = La and Ce) were grown out of a ternary melt with excess As and Cd. 
The constituent elements of high purity La (Ce), Cd, and As, taken in the ratio $R_{0.04}$(Cd$_{0.44}$As$_{0.56}$)$_{0.96}$, were placed in an alumina crucible and sealed in a silica ampoule under a partial Ar pressure. Initially the ampoule was heated slowly to 450~$^\circ$C in a box furnace, the temperature was maintained for 5~hours, and then was finally heated to 950~$^\circ$C over 24 hours. After being held for 1~hour at 950~$^\circ$C, the ampoule was slowly cooled down to 800~$^\circ$C over 100~hours. 
After removing the excess liquid by centrifuging, shiny single crystals were obtained (see insets of Fig.~\ref{FIG1}). Note that when the initial Cd composition is higher than the As composition and the temperature is raised above 1000~$^\circ$C, hexagonal $R$Cd$_{3}$As$_{3}$ phases are stabilized~\cite{Dunsiger2020}. The crystal structure of the samples was confirmed by powder X-ray diffraction (XRD) at room temperature. Figure~\ref{FIG1} shows XRD patterns of powdered LaCd$_{0.67}$As$_{2}$ and single crystal CeCd$_{0.67}$As$_{2}$. No major differences in X-ray patterns are observed between $R$ = La and Ce, except a shift of peak positions due to differences in the unit cell volume.

The peak positions in powder XRD patterns can be well indexed with a monoclinic structure ($C2/m$, no. 12), which is reported to be a vacancy ordered superstructure of the tetragonal SrZnBi$_{2}$-type structure ($I4/mmm$, no. 139)~\cite{Stoyko2011, Piva2021}. 
The obtained unit cell parameters in monoclinic structure are $a = 21.7408$~\AA, $b = 4.0995$~\AA, $c = 12.3220$~\AA, $\alpha = \gamma = 90$$^\circ$, and $\beta = 100.96$$^\circ$ for LaCd$_{0.67}$As$_{2}$ and $a = 21.6126$~\AA, $b = 4.0528$~\AA, $c = 12.2154$~\AA, $\alpha = \gamma = 90$$^\circ$, and $\beta = 100.71$$^\circ$ for CeCd$_{0.67}$As$_{2}$. When the high intensity peaks (main peaks) are only considered, the Bragg reflections of both La- and Ce-based compounds can be indexed in a tetragonal SrZnBi$_{2}$-type structure~\cite{Stoyko2011, Piva2021}. Unit cell parameters refined for the tetragonal SrZnBi$_{2}$-type representation are obtained to be $a = b = 4.1052$~\AA~and $c = 21.3444$~\AA~for $R$ = La and $a = b = 4.0689$~\AA~and $c = 21.2363$~\AA~for $R$ = Ce, which are consistent with previous report~\cite{Piva2021}. The grown single crystals form plates, as shown in the insets of Fig.~\ref{FIG1}, where well formed edge facets are clearly visible. The observation of only ($h$, 0, 0) peaks, as shown in Fig.~\ref{FIG1}(b), confirms that the crystallographic $a$-axis in monoclinic structure is perpendicular to the plane of the plates. The ($h$, 0, 0) reflections in the monoclinic representation correspond to the (0, 0, $l$) reflections in the tetragonal representation. For the physical property measurements we use the tetragonal representation to denote the magnetic field orientations: $H \parallel c$ and $H \perp c$. Note that the magnetic field orientation $H \perp c$ is either $H \parallel$ [100] or [010], otherwise specified.

The dc magnetization was measured in a Quantum Design (QD) Magnetic Property Measurement System (MPMS) up to 70 kOe. The specific heat,  $C_{p}$, was measured by the relaxation technique down to $T$ = 1.8~K in a QD Physical Property Measurement System (PPMS). Four-probe dc resistivity measurements were performed in a QD PPMS with the current flowing perpendicular to the $c$-axis. Due to the high contact resistance, of the order of $\sim$30 $\Omega$ at room temperature, a current of less than 1 $\mu$A was applied below the phase transition ($T_{s}$) to avoid sample heating. Hall resistivity measurements were performed in a four-wire geometry, where the current was flowing normal to the $c$-axis and the magnetic field was applied to the $c$-axis. To remove magnetoresistance effects due to voltage-contact misalignment the magnetic field directions were reversed. It has to be noted that due to the very large longitudinal resistivity contribution caused by the voltage contact misalignment, the Hall resistivity was measured only down to 50 K. Thermoelectric power measurements were performed with two thermometer and one heater configuration in a home-made setup.

\section{Result and discussion}

\begin {figure}
\includegraphics[width=1\linewidth]{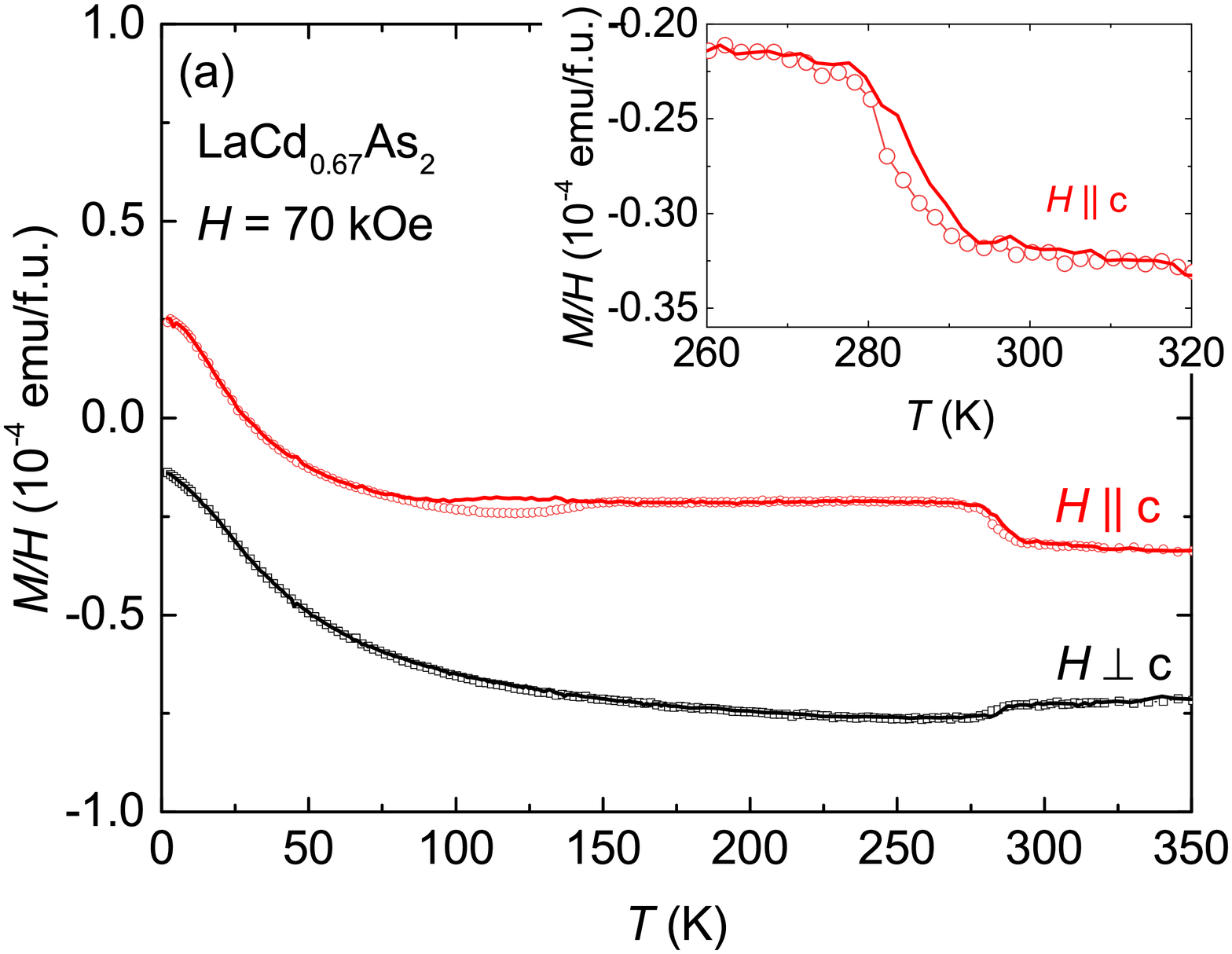}
\includegraphics[width=1\linewidth]{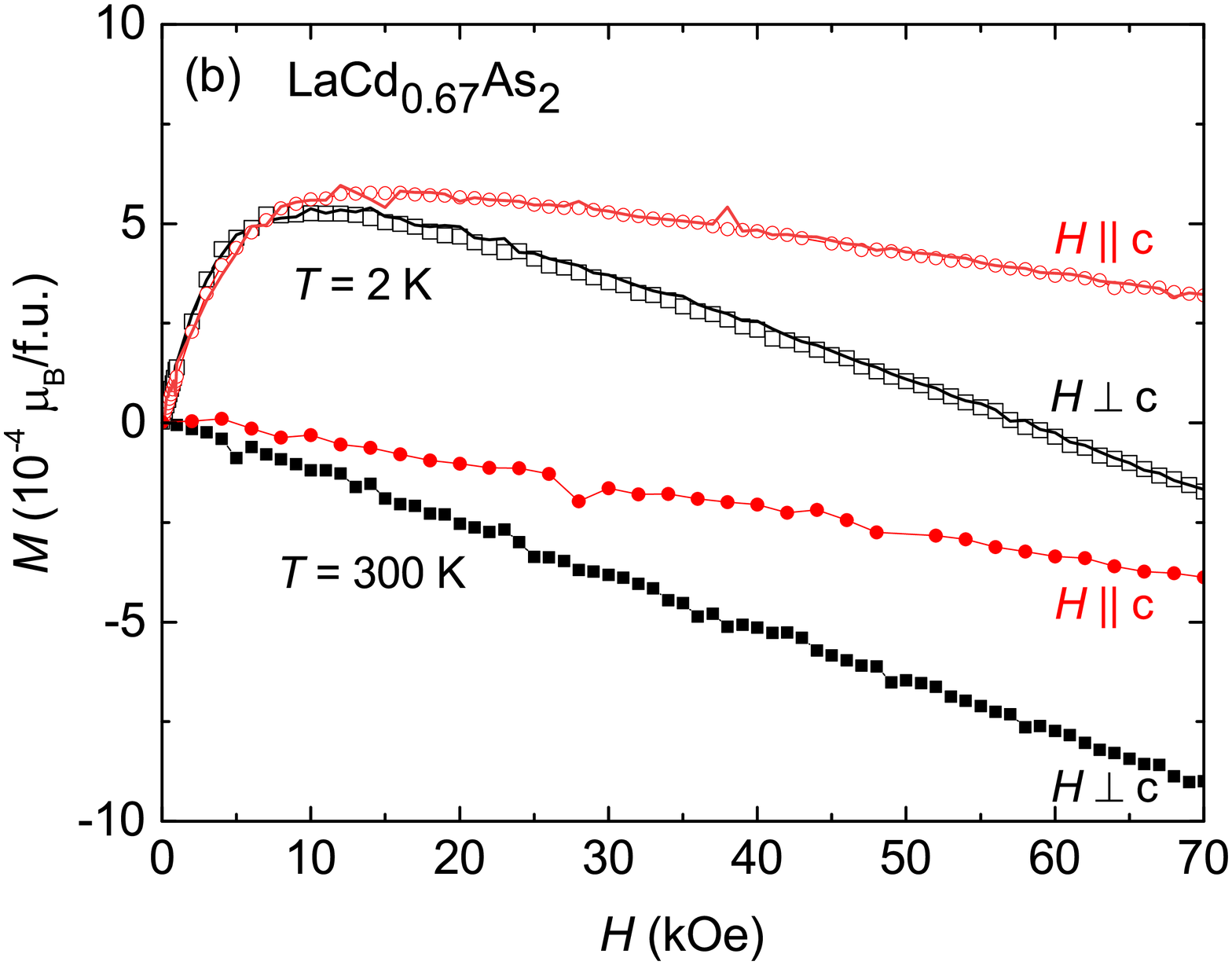}
\caption{(a) $M/H$ of LaCd$_{0.67}$As$_{2}$ for both $H \parallel c$ and $H \perp c$. The magnetic field orientations are based on the tetragonal representation. Symbols and lines indicate the up-sweep and down-sweep of temperature, respectively. Inset shows an enlarged plot between 260 and 320~K for $H \parallel c$. (b) $M(H)$ at 2 and 300~K. Symbols and lines indicate the up-sweep and down-sweep of magnetic field, respectively.}
\label{FIG2}
\end{figure}

Magnetic susceptibility of LaCd$_{0.67}$As$_{2}$ reveals remarkable temperature dependences, with a hint of a (structural) phase transition and a hysteresis. Figure \ref{FIG2}(a) shows anisotropic magnetic susceptibility, $\chi(T) = M/H$, curves below 350~K at $H$ = 70 kOe, applied both perpendicular ($H \perp c$) and parallel ($H \parallel c$) to the $c$-axis in the tetragonal representation of the crystal structure. $M/H$ of LaCd$_{0.67}$As$_{2}$ depends weakly on temperature with a small magnitude, where the absolute value of magnetic susceptibility for $H \perp c$ is roughly twice bigger than that for $H \parallel c$. $M/H$ curves for both orientations indicate a jump with a small hysteresis loop between 280 and 290~K as a signature of a (structural) phase transition (inset). Note that the jumps seen in $M/H$ for two different orientations are in opposite direction, implying anisotropic change of either Landau diamagnetism or Pauli paramagnetism. $M/H$ for $H \parallel c$ clearly indicates a hysteresis between 70 and 150~K while warming and cooling, whereas the hysteresis for $H \perp c$ is negligible.

At low temperatures, $M/H$ curves for both field orientations show an upturn, most likely due to the presence of paramagnetic impurities, consistent with the magnetic field dependence of magnetization at 2~K as shown in Fig.~\ref{FIG2}(b). Magnetization isotherms, $M(H)$, at $T$ = 300~K decrease quasi-linearly with increasing magnetic field up to 70~kOe for both orientations, whereas $M(H)$ curves at $T$ = 2~K show a hump around 10~kOe probably due to the presence of paramagnetic impurities and indicate basically linear field dependences above 20~kOe. It should be noted that the slope in linear field regime at 2~K is the same as that at 300~K. $M(H)$ curves at 2~K show no hysteresis while increasing and decreasing magnetic field for both $H \perp c$ and $H \parallel c$ directions.

\begin {figure}
\includegraphics[width=1\linewidth]{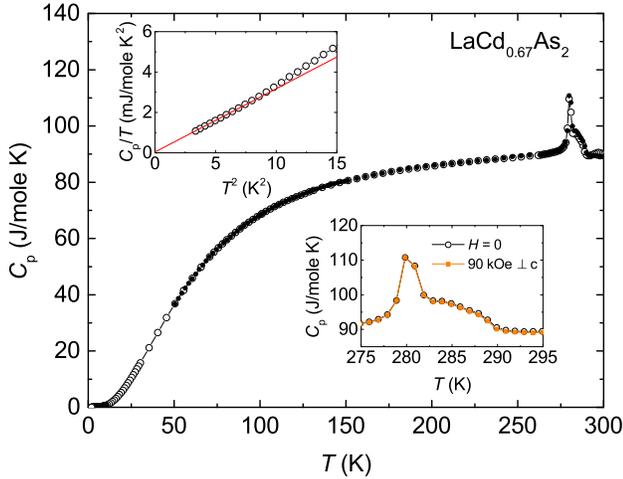}
\caption{$C_{p}$ of LaCd$_{0.67}$As$_{2}$ in zero field. Top inset shows $C_{p}$/$T$ vs $T^2$ plot. The solid line represents the linear fit below 2.5~K. Bottom inset shows an enlarged plot near the phase transition at $H$ = 0 and 90~kOe for $H \perp c$.}
\label{FIG3}
\end{figure}

A jump observed in $M/H$ through $280 \sim 290$~K is clearly confirmed as a phase transition from the specific heat measurements. The temperature dependence of the specific heat, $C_{p} (T)$, curves of LaCd$_{0.67}$As$_{2}$ are shown in Fig. \ref{FIG3}, where $C_{p}$ data are taken for both warming and cooling. A well-defined $\lambda$-shaped anomaly at $\sim$280~K on top of a broad background between 280 and 290~K is clearly seen (inset), confirming the phase transition. This high temperature anomaly shows no magnetic field dependence up to $H = 90$~kOe applied for $H \parallel c$, as shown in the bottom inset. The phase transition temperature, corresponding to the sharp peak position at $T_{s}$ = 280 K, is consistent with earlier report \cite{Piva2021}. However, the broad maximum underneath the peak observed in this study (flux grown sample) was not identified in earlier study on La$_{3}$Cd$_{2}$As$_{6}$ sample grown by the CVT technique \cite{Piva2021}. Although the origin of the broad background is currently unclear, we conjecture that this is related to the distortion of As-square net \cite{tremel1987} and the sample inhomogeneity.

Above $T_{s}$ the absolute value of $C_{p}$ reaches $\sim$90 J/mole K, which is somewhat lower than the Dulong-Petit limit. 
Unlike the magnetic susceptibility for $H \parallel c$, $C_{p}$ indicates no hysteresis between 70 and 150~K. Since the specific heat curve follows $C_{p}(T) = \gamma T + \beta T^{3}$ in a limited temperature range (top inset), the electronic and phonon contributions are estimated by fitting the curve below 2.5~K. The electronic contribution and Debye temperature are estimated to be $\gamma \sim$ 0.036 mJ/mole K$^2$ and $\Theta_{D}$ $\sim$ 283~K, respectively, which are consistent with earlier study \cite{Piva2021}. The small $\gamma$ value suggests either a very small effective mass or low carrier density in this material.

\begin {figure}
\includegraphics[width=1\linewidth]{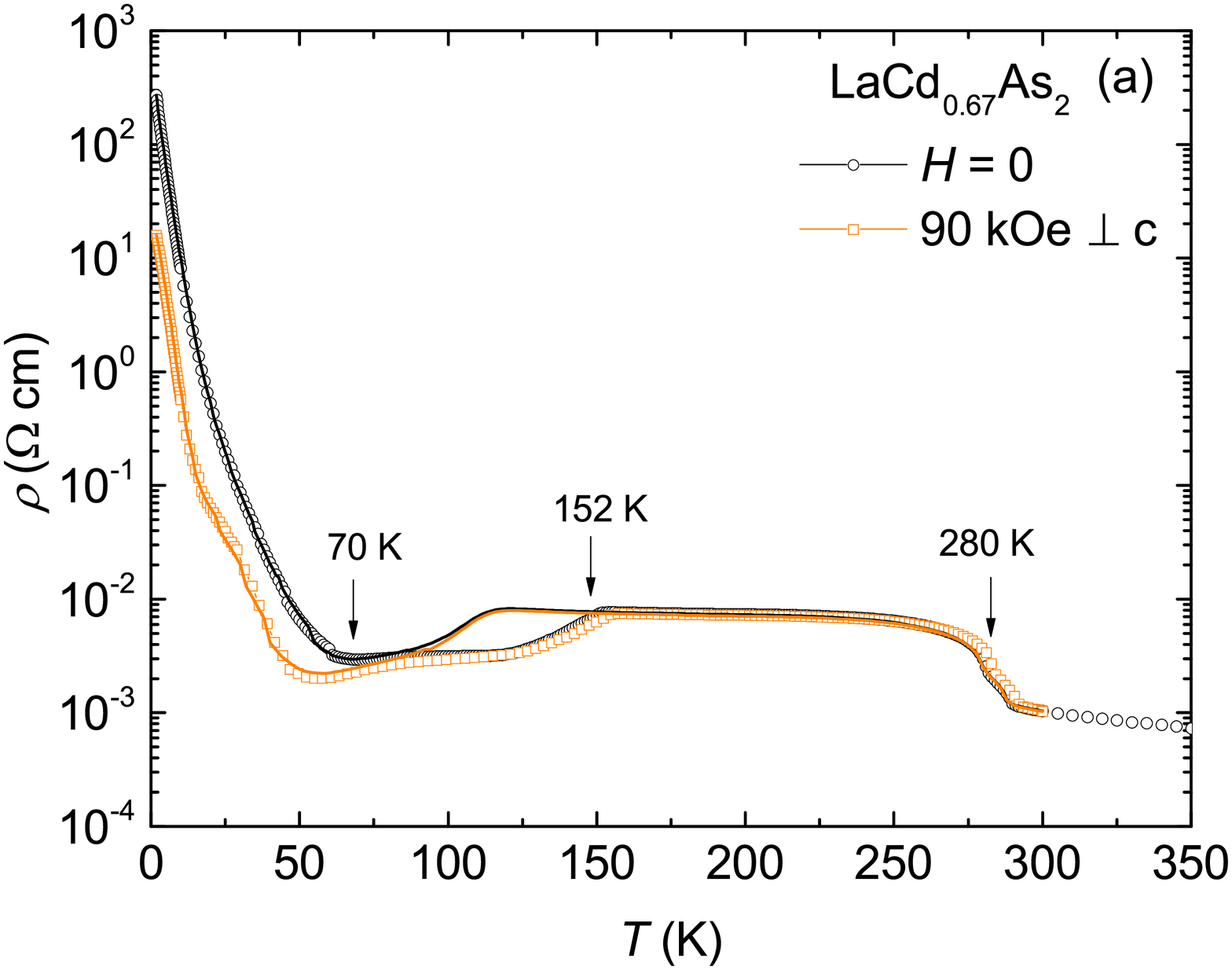}
\includegraphics[width=1\linewidth]{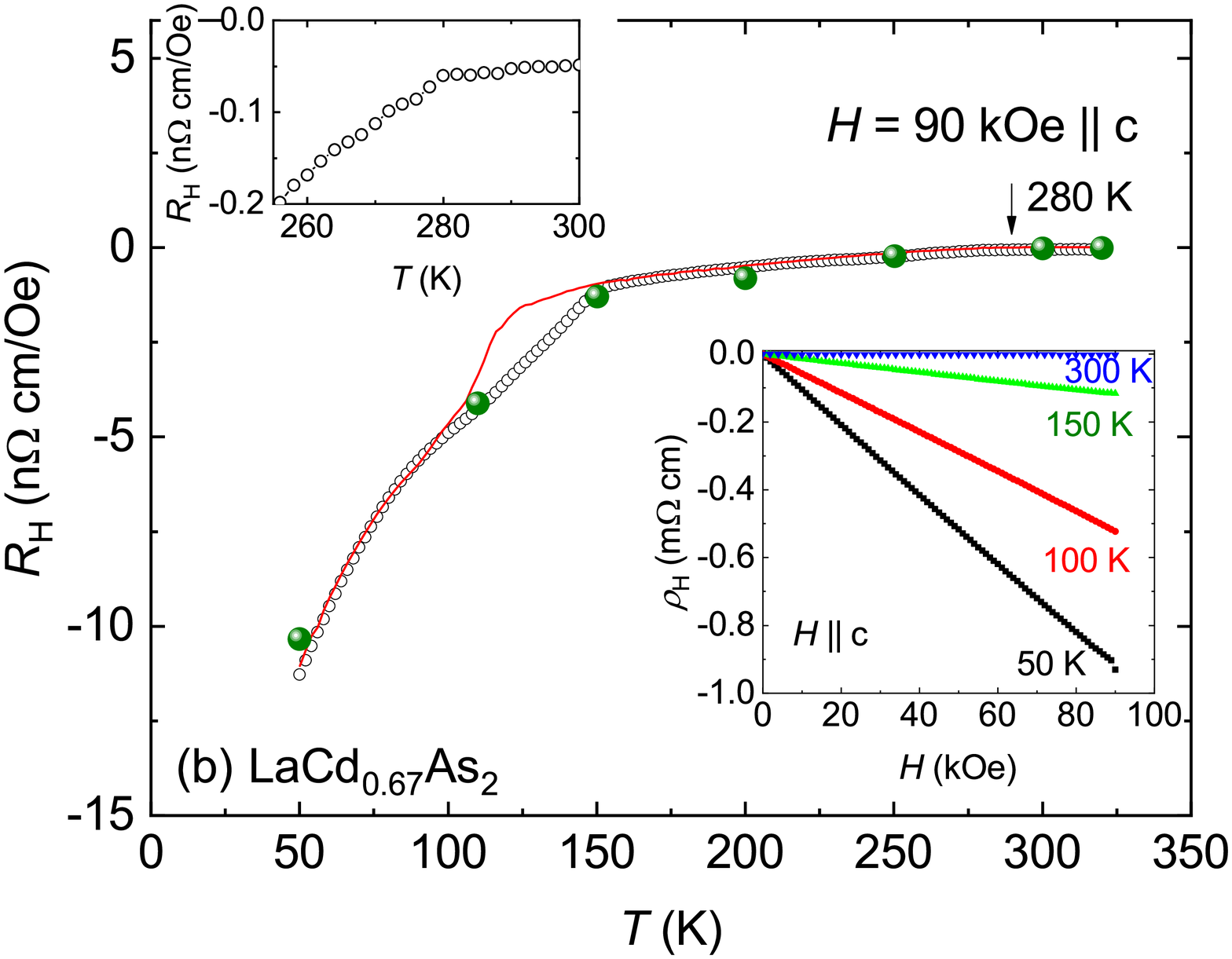}
\includegraphics[width=1\linewidth]{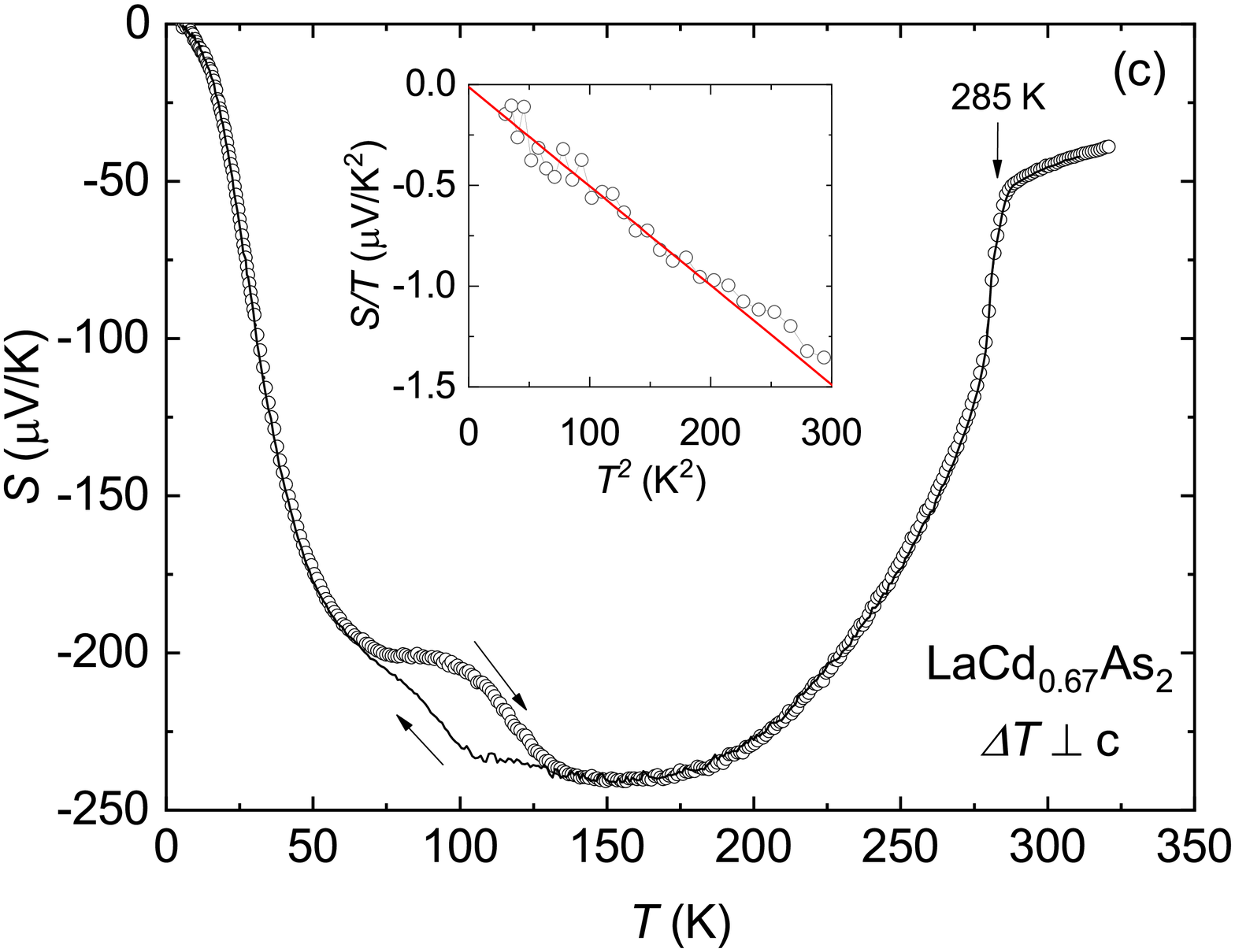}
\caption {(a) $\rho(T)$ of LaCd$_{0.67}$As$_{2}$ at $H$ = 0 and 90~kOe for $H \perp c$. (b) $R_H$ at 90 kOe for $H \parallel c$. Open symbol and line indicates $R_H$ taken while warming and cooling, respectively. Solid symbols are taken from the magnetic field sweep while cooling the sample. Top inset shows $R_H$ near 280~K. Bottom inset shows Hall resistivity, $\rho_{H}$, at selected temperatures. (c) Zero field $S(T)$, taken while warming and cooling. Inset shows $S/T$ vs $T^{2}$ plot.}
\label{FIG4}
\end{figure}

The temperature dependence of the electrical resistivity, $\rho(T)$, curves of LaCd$_{0.67}$As$_{2}$ are shown in Fig.~\ref{FIG4}(a) at $H$ = 0 and 90~kOe for $H \perp c$, where the resistivity measurements are performed while warming and cooling. In zero field, $\rho(T)$ indicates three distinctive behaviors: a jump below 290~K with minor hysteresis, a pronounce hysteresis between 70 and 152~K, and a semiconductor-like increase below 70~K. $\rho(T)$ at 90 kOe shows a negative magnetoresistance below 70~K, while no magnetic field dependence is shown above 70~K. It has to be noted that $\rho(T)$ of CVT grown La$_{3}$Cd$_{2}$As$_{6}$ indicates a remarkable increase of thirteen order of magnitude below the phase transition at 279~K \cite{Piva2021}, which is in sharp contrast to the flux grown sample.

A large resistivity value and rapid increase of resistivity below 70~K suggest a low carrier concentration in LaCd$_{0.67}$As$_{2}$. To estimate the carrier concentration, the Hall resistivity, $\rho_{H}$, is measured as a function of temperature at 90~kOe for $H \parallel c$. The temperature dependence of the Hall coefficient, $R_{H} = \rho_{H}/H$, is plotted in Fig.~\ref{FIG4}(b). The open symbol and solid line indicates $R_{H}$ measurements while warming and cooling the temperature, respectively. As the temperature decreases $R_{H}$ curve shows a very weak slope change at $\sim$280~K, reveals a hysteresis loop between 90 and 152~K, and drastically decreases below the hysteresis, which are consistent with $\rho(T)$ behavior. The sharp change in $R_H$ can be related to the decreasing carrier density. Based on the one-band model approximation, the carrier density is estimated to be $\sim$1.29$\times10^{21}$ per cm$^3$ at 300~K and $\sim$5.5$\times10^{18}$ per cm$^3$ at 50 K, which corresponds to $\sim$0.12 carrier per formula unit (f.u.) at 300~K and $\sim$0.0005 carrier per f.u. at 50~K, confirming the low carrier density material. Thus, the negligibly small $\gamma$ value can be related to the low carrier density in LaCd$_{0.67}$As$_{2}$. The negative sign of $R_H$ indicates that transport is dominated by electron-like carriers, which is consistent with thermoelectric power (TEP) measurement.

As a further probe of the transport property, TEP of LaCd$_{0.67}$As$_{2}$ has been measured as a function of temperature, $S(T)$, plotted in Fig. \ref{FIG4}(c). $S(T)$ indicates a sharp drop below 285~K and shows a parabolic temperature dependence with a pronounced hysteresis loop between 70 and 132~K. The minimum in $S(T)$ occurs near 150~K with $-240$~$\mu$V/K. The observed TEP value of LaCd$_{0.67}$As$_{2}$ is an order of magnitude higher than the one in typical metals and close to the typical values of narrow-gap semiconductors~\cite{Dughaish2002, Zachary2015, Witting2019}. Below 150~K the rapid variation in $S(T)$ can be related the thermal depopulation of charge carriers. In a multi band system, the Seebeck coefficient is given by the Seebeck coefficient of each band weighted by the electrical conductivity of each band~\cite{Behnia2004}. Therefore, to explain the strong temperature dependence of TEP both carrier density and mobility of each band must be properly accounted.

At low temperatures, $S(T)$ can be very well described by $a T + b T^{3}$ relation, as shown in the inset, where the first and second term represent the electronic and phonon contribution, respectively. By fitting the curve below 10~K (inset) the electronic contribution is estimated to be $a = -1.28 \times 10^{-8}$ V/K$^{2}$. The small value of $a$ yields the Lorentz ration $q = a N_{A}e/\gamma$ $\sim - 34$, where the constant $N_{A}e = 9.6 \times 10^{4}$ C/mole is the Faraday number~\cite{Behnia2004}. The obtained large $q$ value implies a low carrier density of the LaCd$_{0.67}$As$_{2}$ system, which is consistent with Hall effect measurements. In general, $S(T)$ of typical metals shows a maximum, corresponding to the phonon-drag effect, where the maximum is expected to be located between $\Theta_{D}/5$ and $\Theta_{D}/12$~\cite{Blatt1976}. $S(T)$ of LaCd$_{0.67}$As$_{2}$ indicates an absence of the low temperature broad peak caused by the phonon drag.

Thermodynamic and transport property measurements of LaCd$_{0.67}$As$_{2}$ clearly indicate the phase transition below 280~K. It has been shown that the distortion of As- and Sb-square net in this family of materials induces a metal to semiconductor transition \cite{tremel1987, Rutzinger2010, Stoyko2011}. Since LaCd$_{0.67}$As$_2$ compound has no moment bearing ions, the anomaly can't be related to magnetic origin, but can be associated with electronic structure change and/or structural phase transition. In CVT grown La$_{3}$Cd$_{2}$As$_{6}$ sample, a sharp phase transition at 279~K has been attributed to a charge density wave (CDW) transition caused by a distortion in the arsenic square net \cite{Piva2021}. Among these family of materials, CDW transition has been identified from exceptionally clean LaAgSb$_{2}$ single crystal \cite{C.song2003}. We tentatively anticipate that the discrepancy between CVT and flux grown samples could be associated with structural features like different level of local atomic distortions and Cd site vacancies. In addition, we conjecture that additional features observed below $T_{s}$ in transport properties of LaCd$_{0.67}$As$_{2}$ are related to further structural distortions and superstructure that may change the electronic properties over the temperature range measured. In the superstructure of $RT$As$_{2}$ ($T$ = Zn and Cd) the vacancy ordering at the zinc site of the primitive tetragonal HfCuSi$_{2}$-type structure forms an orthorhombic ($Pmmn$) superstructure (Pr$_{3}$Zn$_{2}$As$_{6}$) \cite{Nientiedt1999}, whereas the Cd vacancy ordering of the body centered tetragonal ZrZnBi$_2$-type structure forms a monoclinic ($C2/m$) superstructure ($R_{3}$Cd$_{2}$As$_{6}$) \cite{Piva2021}. Therefore, it is necessary to perform high energy X-ray experiments to understand the observed physical properties of the flux grown LaCd$_{0.67}$As$_{2}$ sample.

\begin {figure}
\includegraphics[width=1\linewidth]{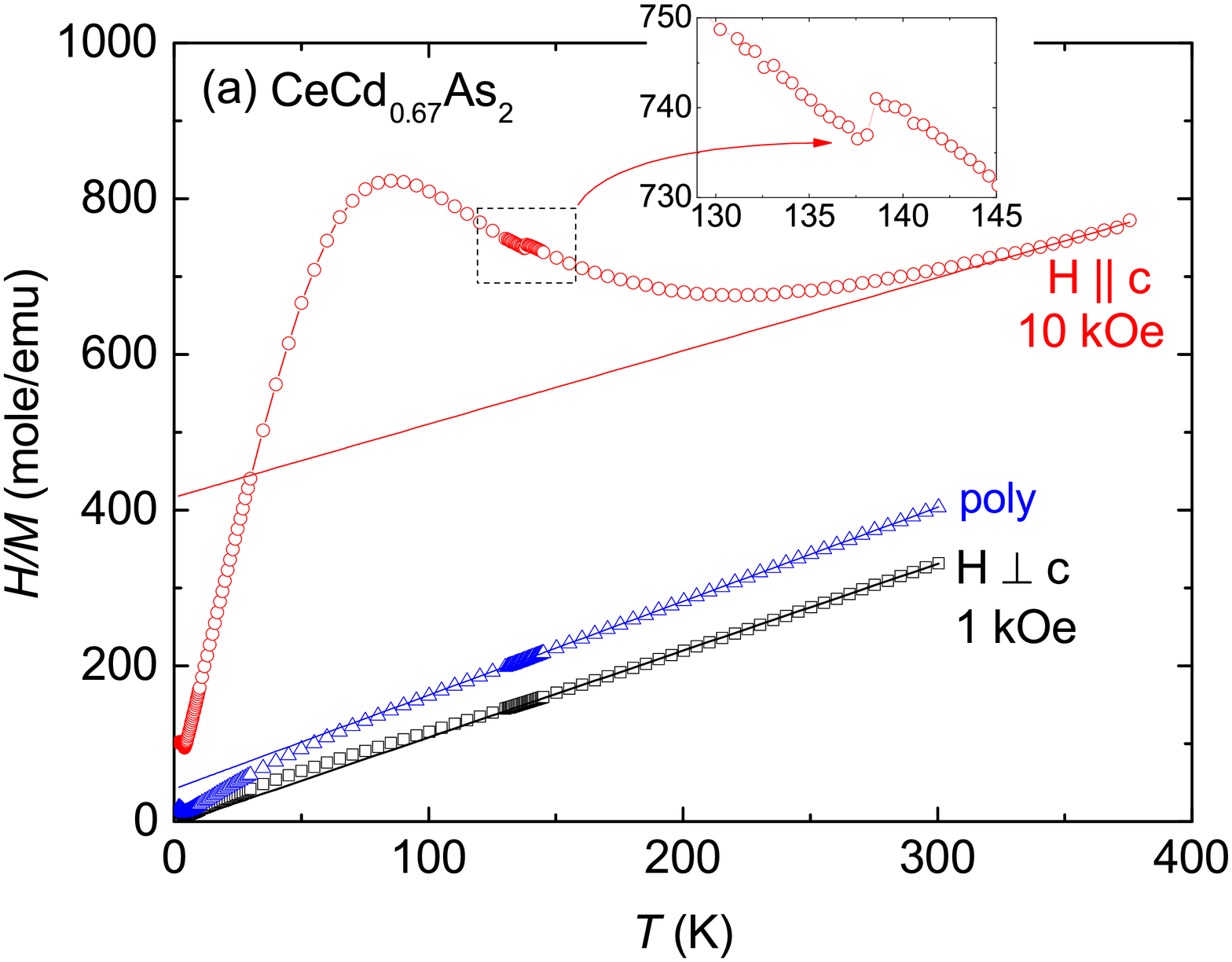}
\includegraphics[width=1\linewidth]{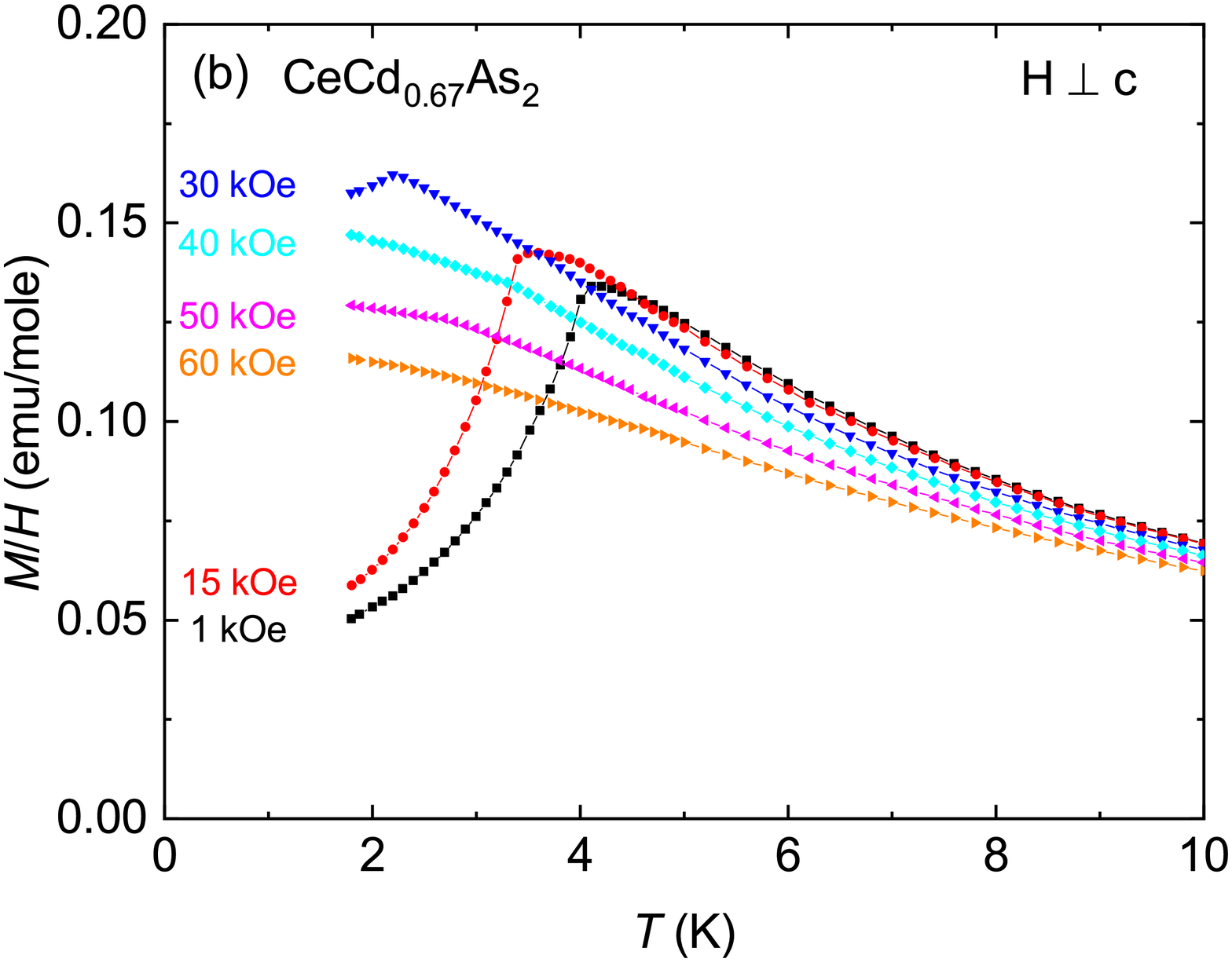}
\includegraphics[width=1\linewidth]{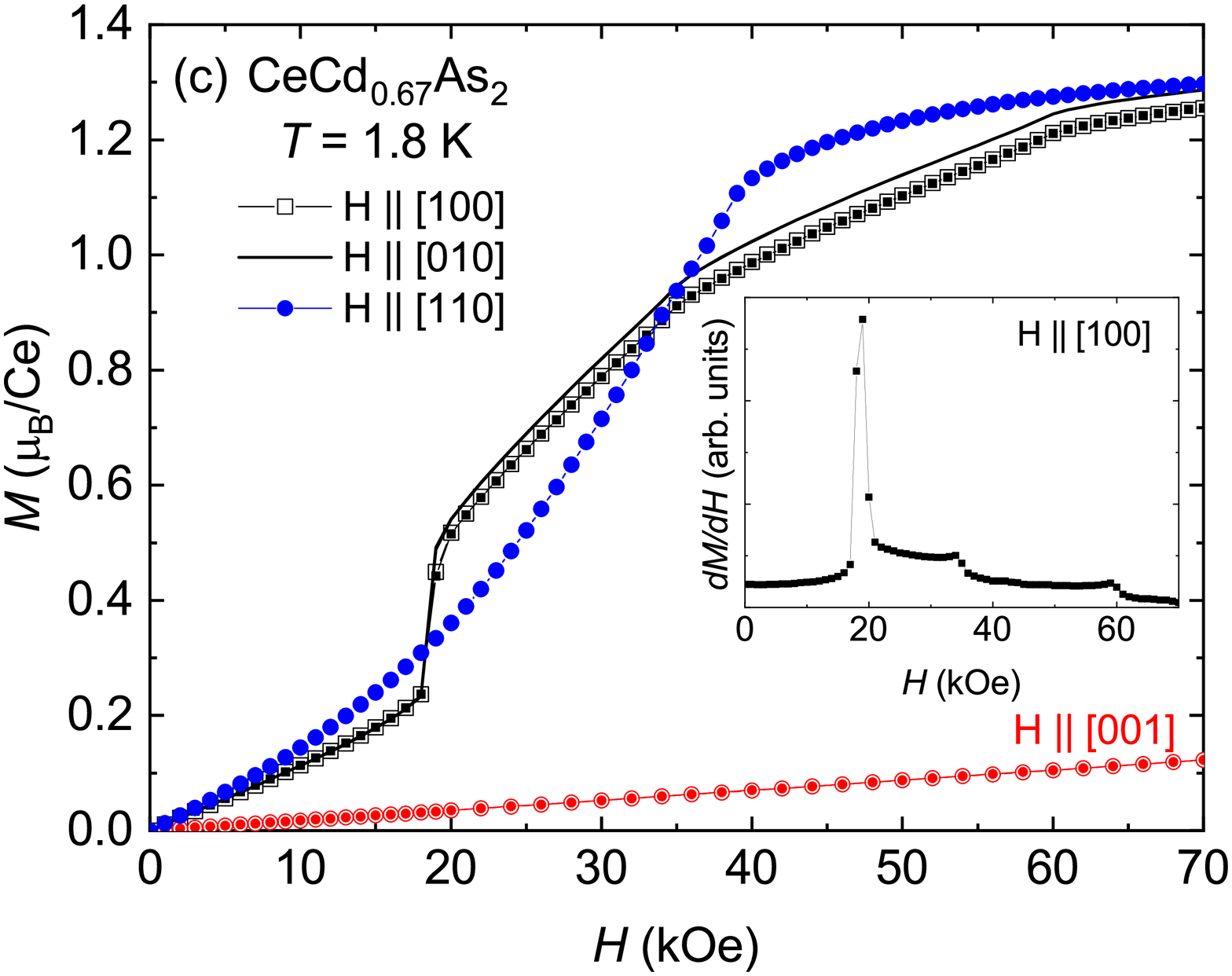}
\caption {(a) Inverse magnetic susceptibility, $H/M$, of CeCd$_{0.67}$As$_{2}$ for $H \perp c$, $H \parallel c$, and poly average crystalline. Solid lines represent C-W fit. (b) $M/H$ of CeCd$_{0.67}$As$_2$ for $H \perp c$ at selected magnetic fields. (c) Magnetization isotherms at 1.8~K for $H \parallel$ [100], [010], [110], and $H \parallel$ [001] in the tetragonal representation. Inset shows $dM/dH$ for $H \parallel$ [100].}
\label{FIG5}
\end{figure}

Inverse magnetic susceptibility, $1/\chi(T) = H/M$, curves of CeCd$_{0.67}$As$_{2}$ are shown in Fig. \ref{FIG5}(a), applied magnetic fields for both $H \perp c$ and $H \parallel c$ and a polycrystalline average in the tetragonal representation. The polycrystalline average is estimated as $\chi_{poly} = 2/3~\chi_{\perp c} + 1/3~\chi_{c}$. It should be emphasized that although the signature is not pronounced, magnetic susceptibility curves, especially for $H \parallel c$, clearly indicates a jump at $T_{s}$ $\sim$137~K as a signature of the phase transition (see inset). $H/M$ for $H \perp c$ follows a linear temperature dependence above 200~K, whereas $H/M$ for $H \parallel c$ shows no clear, linear temperature dependence up to 375~K probably due to the CEF effects. To estimate the effective moment, $\mu_{eff}$, and Weiss temperature, $\theta_{p}$, we fit the curve by the Curie-Weiss law, $\chi(T) = C / (T - \theta_{p})$, from the linear region of the $H/M$ curves: $\mu_{eff}^{\perp c} = 2.55~\mu_{B}$ and $\theta_{p}^{\perp c} = 5.3$~K for $H \perp c$; $\mu_{eff}^{c} = 2.92~\mu_{B}$ and $\theta_{p}^{c} = -442.7$~K for $H \parallel {c}$; and $\mu_{eff}^{poly} = 2.58~\mu_{B}$ and $\theta_{p}^{poly} = -34.7$~K for $\chi_{poly}$. Note that for $H \parallel c$ the fit was performed within the limited temperature range from 350 to 375~K. It is necessary to measure magnetic susceptibility above 375~K to determine $\mu_{eff}^{c}$ and $\theta_{p}^{c}$ more reliably along the $c$-axis. The magnetic susceptibility results in this study are consistent with the earlier report \cite{Piva2022}.

At low temperatures, $\chi(T)$ clearly indicates an antiferromagnetic (AFM) ordering below 4~K. The low temperature $\chi(T)$ data, measured for $H \perp c$, are plotted in Fig.~\ref{FIG5}(b) for various applied magnetic fields. The AFM ordering temperature $T_{N}$, determined from the $d\chi T/dT$ analysis, is consistent with the earlier study \cite{Piva2022}. With increasing magnetic field, $T_{N}$ shifts to lower temperatures and suppresses below 1.8~K for $H > 60$~kOe. In the derivatives, two slope changes appear above 10~kOe and only one peak shows above 35~kOe. Above 60~kOe, $\chi(T)$ does not reveal any signature of phase transitions and instead shows a tendency toward saturation at low temperatures. Note that for $H \parallel c$, $T_{N}$ is hardly affected by the applied magnetic field.

The $M(H)$ measurements, shown in Fig.~\ref{FIG5}(c), shed further light on the low-temperature magnetic states of CeCd$_{0.67}$As$_{2}$. The $M(H)$ curves at 1.8~K are highly anisotropic between $H \perp c$ ($H \perp$ [001]) and $H \parallel c$ ($H \parallel$ [001]). It is clear from the figure that in-plane anisotropy is small. $M(H)$ for $H \parallel$ [001] increases almost linearly up to $\sim$0.13 $\mu_{B}$/Ce$^{3+}$ at 70 kOe. On the other hand, $M(H)$ for $H \parallel$ [100] linearly increases up to 19 kOe and then, undergoes three metamagnetic (MM) transitions and appears to be approaching saturation at 70 kOe with a moment of $\sim$1.25 $\mu_{B}$/Ce$^{3+}$. The MM transitions at $H$ = 19, 34 and 58 kOe are clearly indicated in the $dM/dH$ analysis as shown in the inset. $M(H)$ for both $H \parallel$ [100] and [010] clearly indicates three MM transitions, whereas only two MM transitions are observed for $H \parallel$ [110]. It has to be noted that in the earlier study \cite{Piva2022} two MM transitions at 19 and 34~kOe were only identified in $M(H)$ curve for $H \perp c$ at 2~K. The magnetization value at 70~kOe is consistent with the earlier report \cite{Piva2022} but lower than the theoretical value of 2.14 $\mu_{B}$/Ce$^{3+}$ for the saturated moment of free Ce$^{3+}$ ions. Interestingly, no hysteresis is observed through the MM transitions.

\begin {figure}
\includegraphics[width=1\linewidth]{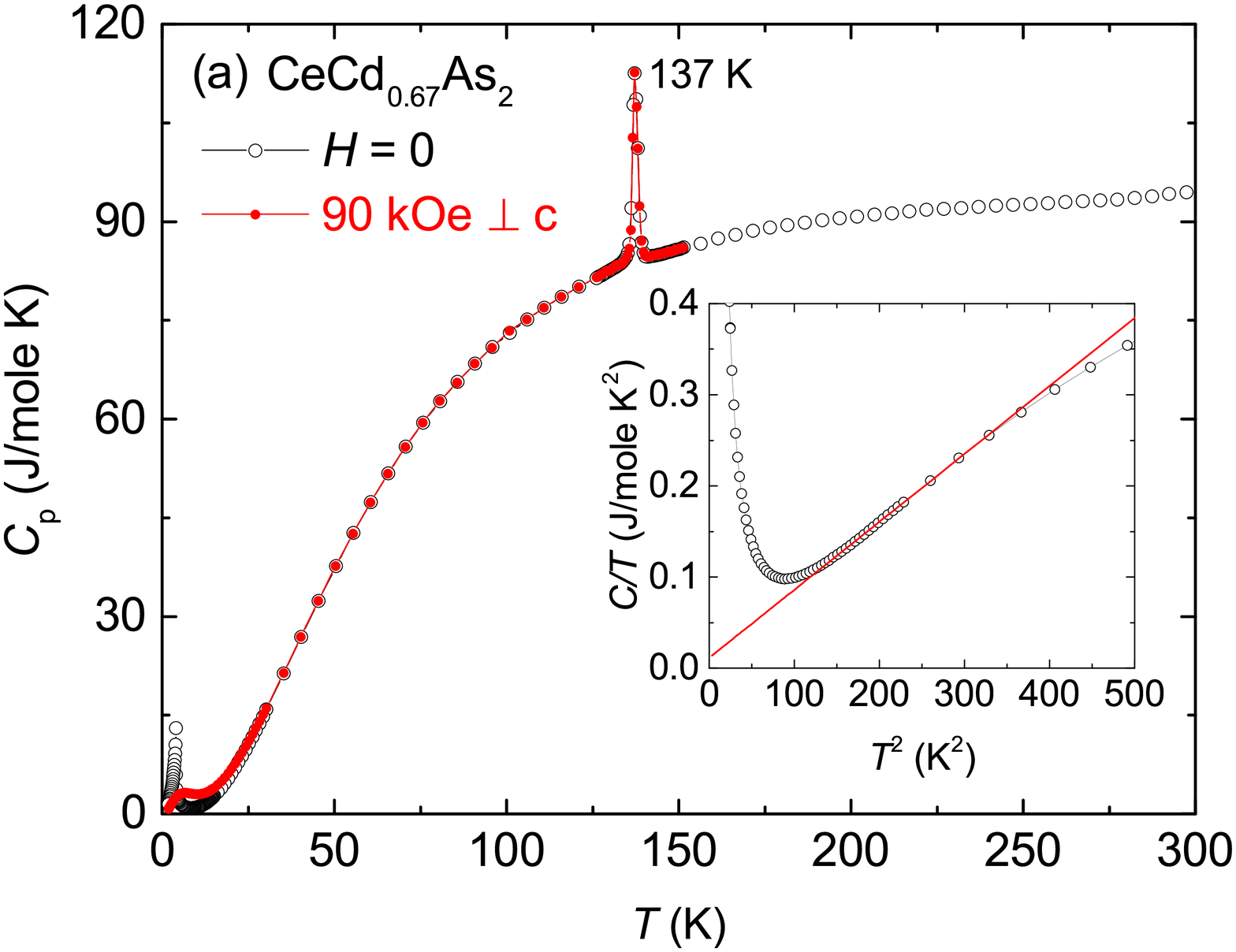}
\includegraphics[width=1\linewidth]{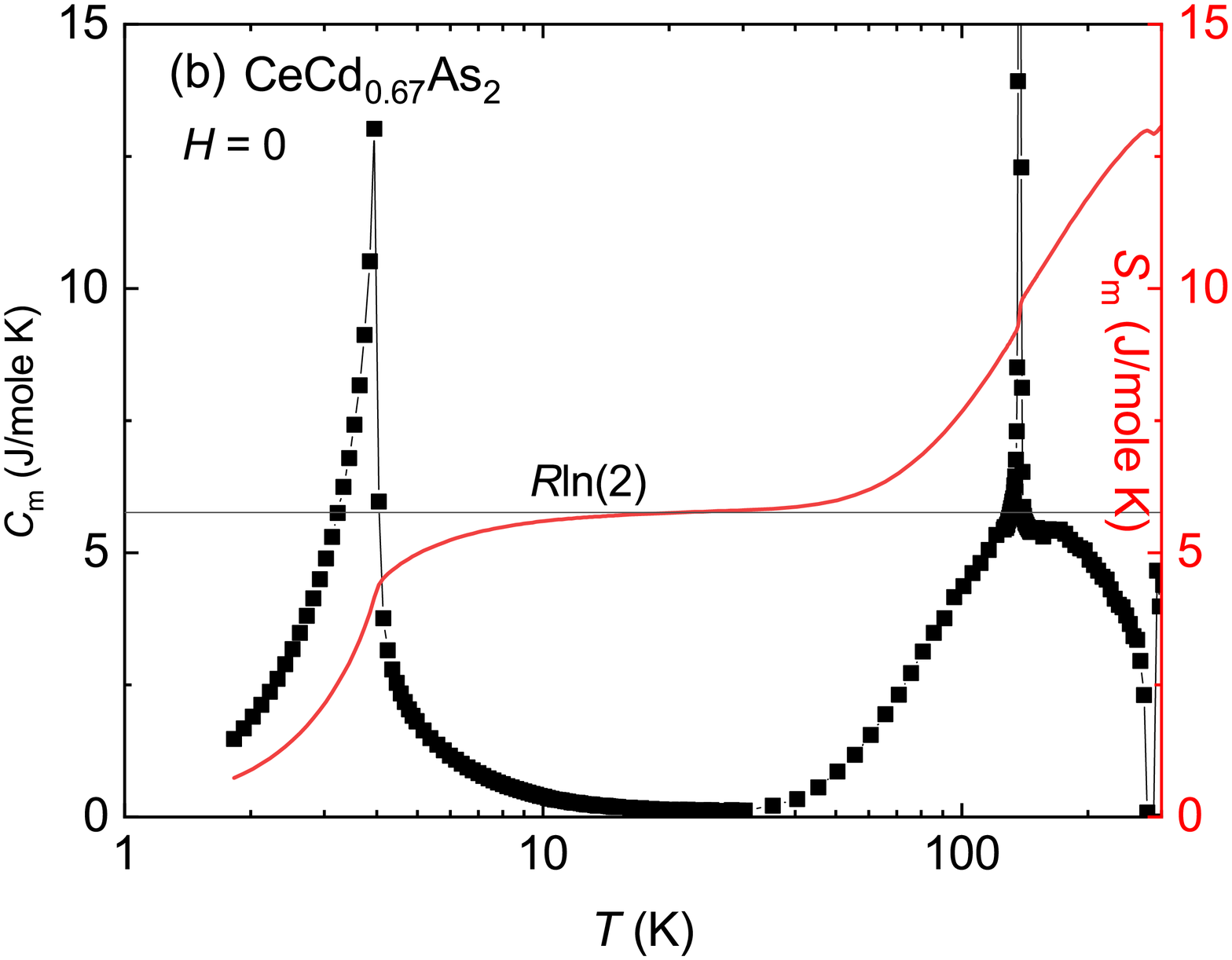}
\includegraphics[width=1\linewidth]{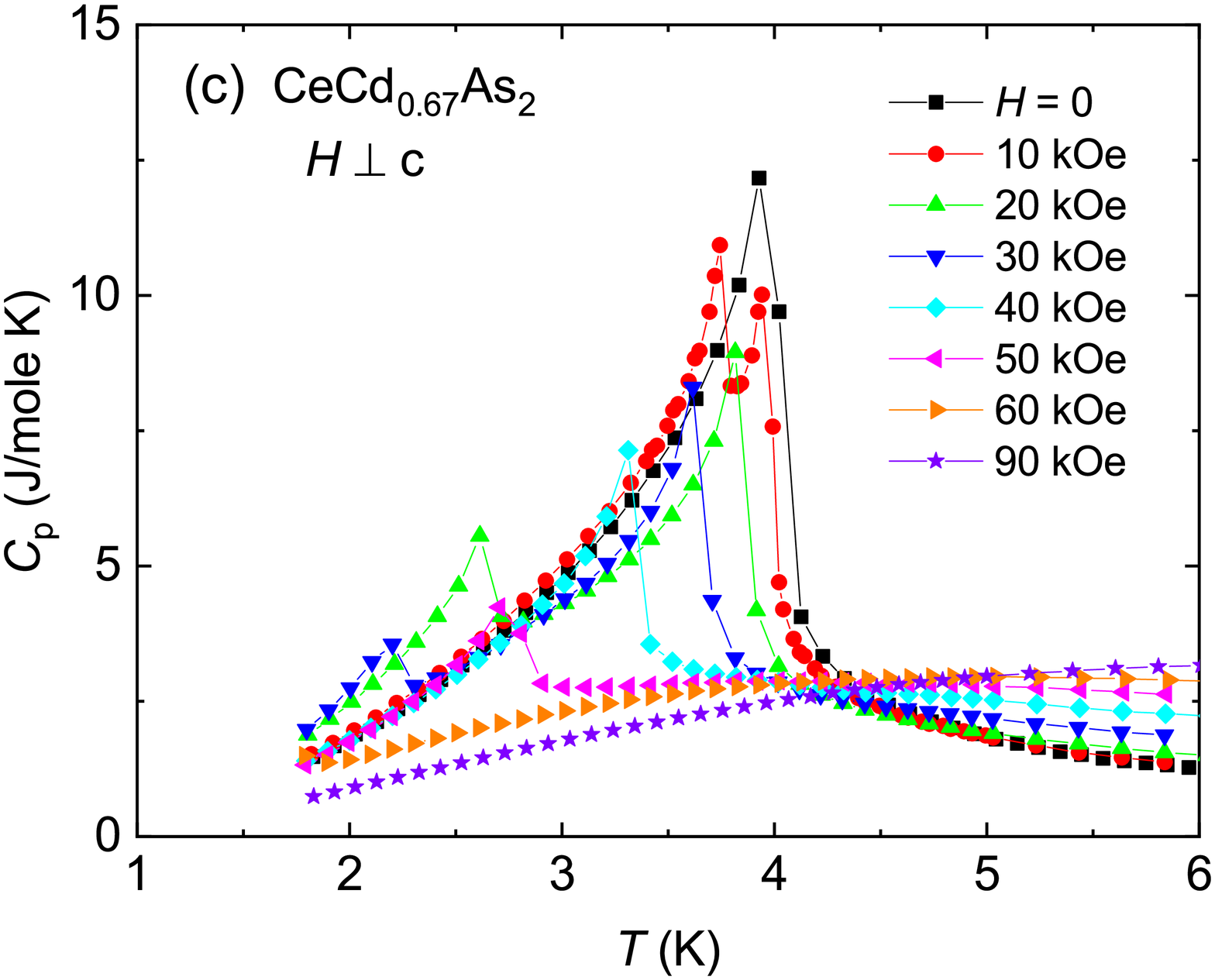}
\caption {(a) $C_{p}$ of CeCd$_{0.67}$As$_{2}$ at $H$ = 0 (open symbol) and 90~kOe (solid symbol) for $H \perp c$. Inset shows $C/T$ vs $T^{2}$ plot. (b) $C_{m}$ (symbol) and $S_{m}$ (line) of CeCd$_{0.67}$As$_{2}$. (c) $C_{p}(T)$ below 6~K at selected magnetic fields.}
\label{FIG6}
\end{figure}

The specific heat of CeCd$_{0.67}$As$_2$, plotted in Fig.~\ref{FIG6}(a), manifests $\lambda$-shape anomalies at 137~K and 4~K, confirming phase transitions observed in $M/H$. Because of the AFM order below 4 K, $\gamma$ and $\Theta_{D}$ of CeCd$_{0.67}$As$_2$ cannot be reliably obtained by using the relation $\gamma T + \beta T^{3}$. Instead, $\gamma$ = 13 mJ/mole $K^{2}$ and $\Theta_{D}$ = 212.7 are estimated by fitting the curve above 10~K, as shown in the inset. The magnetic part of the specific heat, $C_{m}$, is plotted in Fig.~\ref{FIG6}(b), where $C_{p}$ of LaCd$_{0.67}$As$_2$ is used to estimate the nonmagnetic contribution to the specific heat of CeCd$_{0.67}$As$_{2}$. $C_{m}$ shows a broad maximum centered around $\sim$150 K, associated with the CEF splitting of the Hund's rule ground state multiplet.

The magnetic entropy, $S_{m}$, is inferred by integrating $C_{m}/T$ and plotted in Fig.~\ref{FIG6}(b). $S_{m}$ reaches about 80\% of $R\ln(2)$ at $T_{N}$ and recovers the full doublet entropy by 20 K. This suggest that the sharp anomaly at 4 K in specific heat stems from the AFM order in a doublet ground state. In addition, the observed entropy implies that the ordered moment is not compensated by the Kondo interaction. At 300 K the recovered entropy is less than $R\ln(6)$, which suggests a large energy level splitting due to the strong CEF effect.

In order to examine the effect of an applied magnetic field on the AFM order, the specific heat was measured for $H \perp c$ up to 90 kOe, plotted in Fig.~\ref{FIG6}(c). As magnetic field increases, $T_{N}$ shifts to lower temperatures and a second peak develops below $T_{N}$. For $H$ = 60 kOe the phase transition is suppressed and instead $C_{p}$ exhibits a broad maximum centered at 5~K, which marks a crossover from the AFM to paramagnetic state. At higher fields, this maximum broadens further and moves to higher temperatures.

\begin {figure}
\includegraphics[width=1\linewidth]{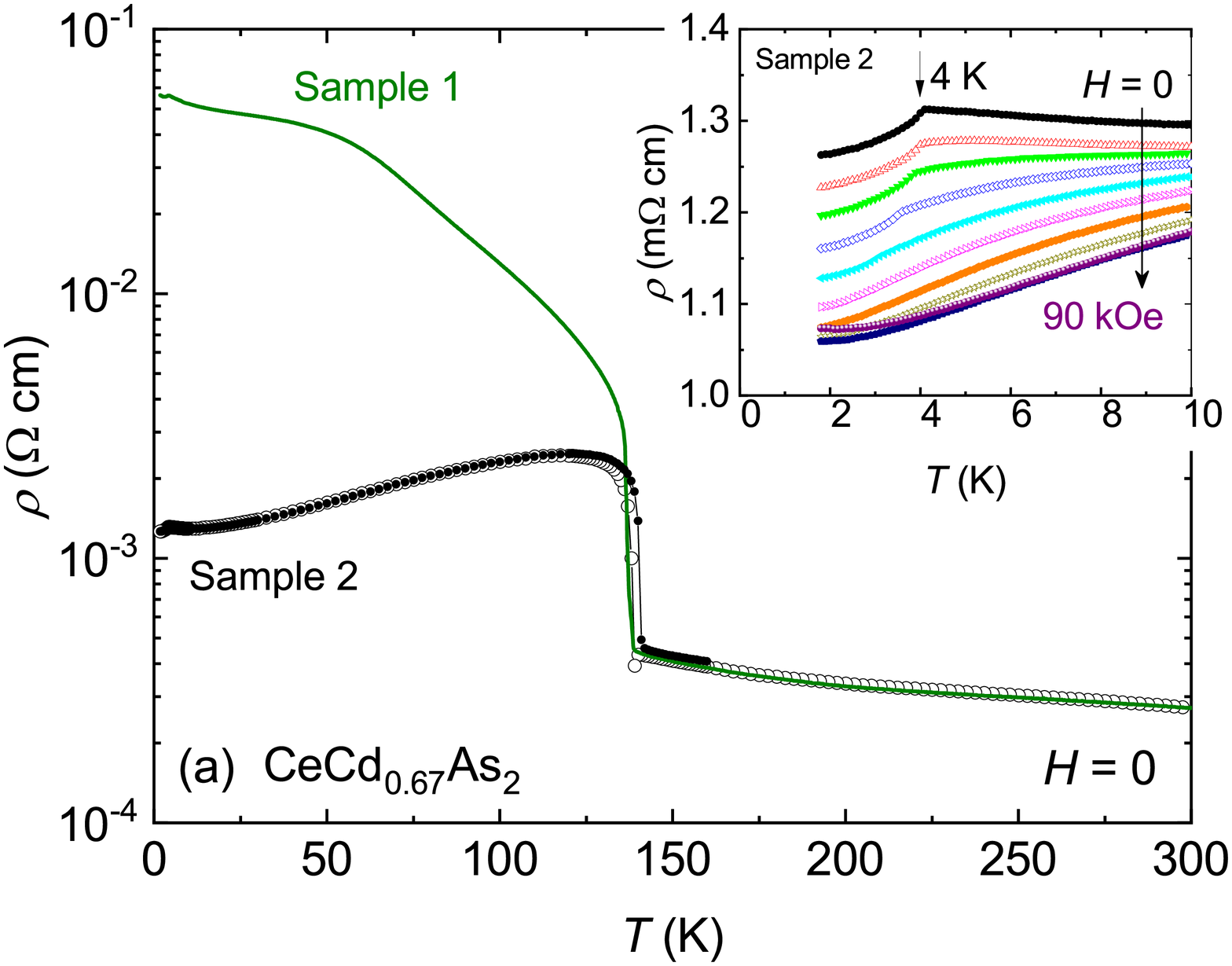}
\includegraphics[width=1\linewidth]{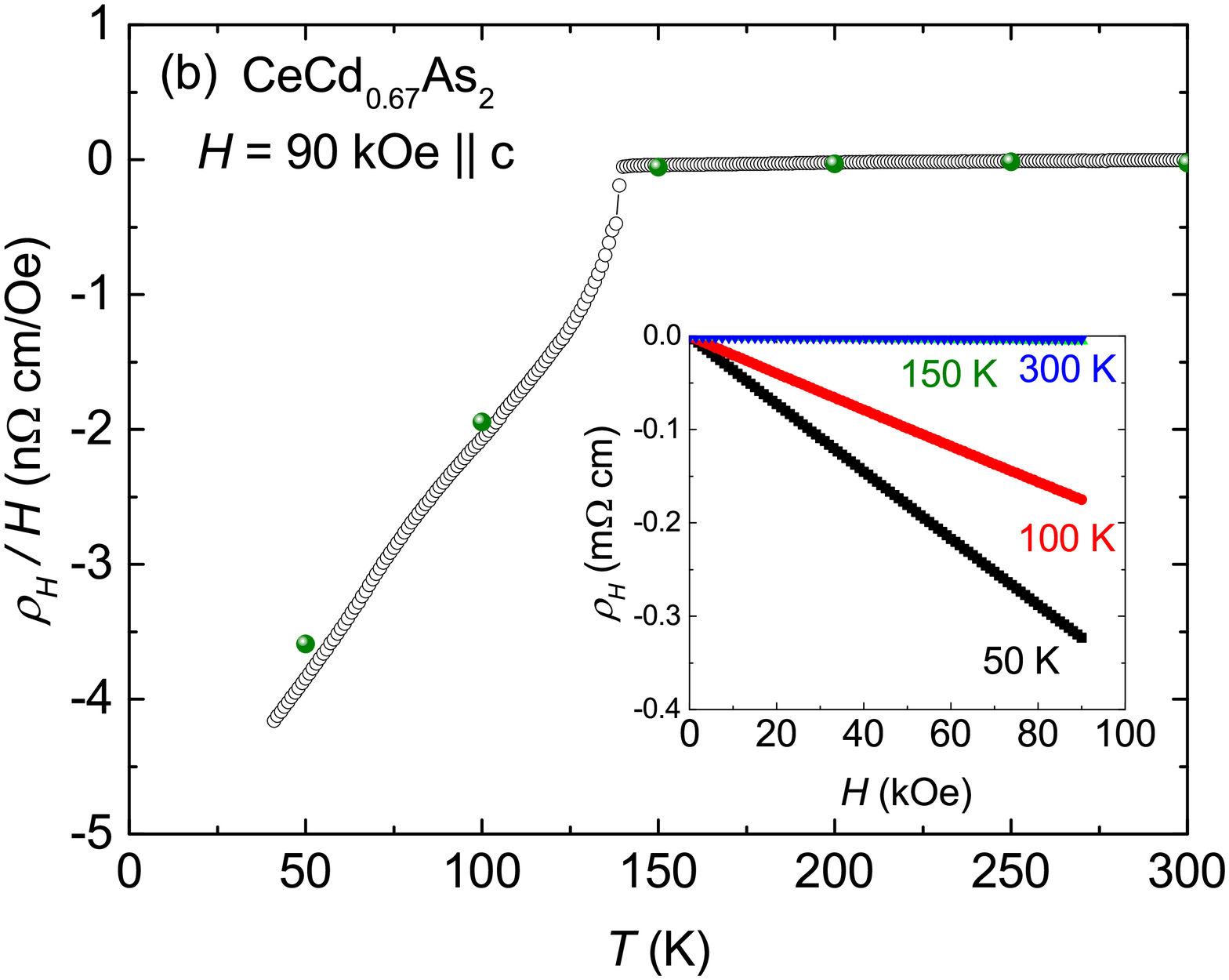}
\includegraphics[width=1\linewidth]{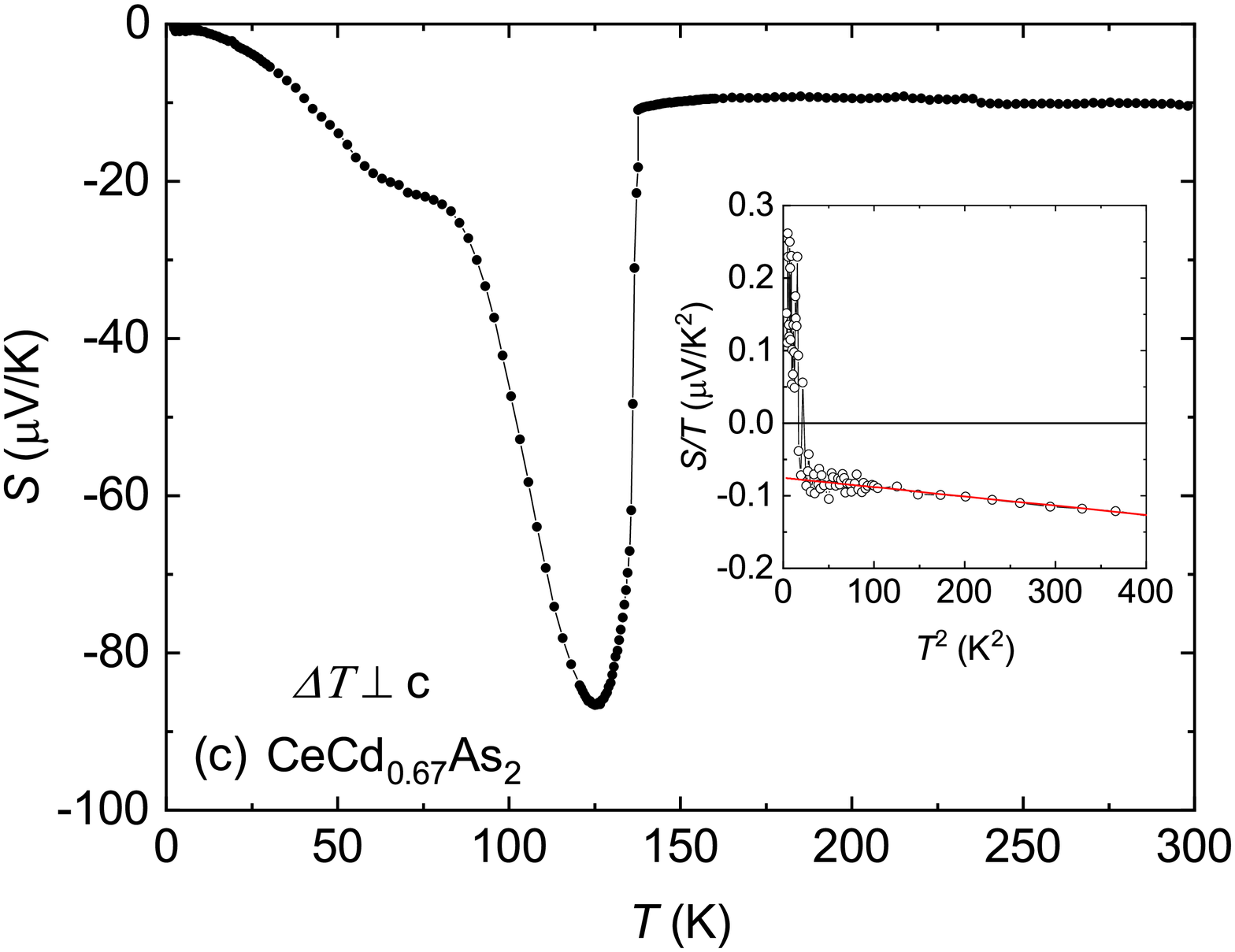}
\caption {(a) $\rho(T)$ of CeCd$_{0.67}$As$_{2}$ for sample 1 and 2. For the sample 2, open and closed symbols represent the resistivity data taken while cooling and warming the temperature. Inset shows $\rho (T)$ of sample 2 at various magnetic fields. (b) $R_H$ at 90 kOe for $H \parallel c$. Solid symbols are taken from the magnetic field sweep data. Inset shows $\rho_{H}$ at selected temperatures. (c) Zero field $S(T)$, taken while warming the sample. Inset shows $S/T$ vs $T^{2}$ plot.}
\label{FIG7}
\end{figure}

The $\rho(T)$ curves of CeCd$_{0.67}$As$_{2}$ are presented in Fig. \ref{FIG7}(a), where two pieces of single crystals (sample 1 and 2) from the same batch are selected to exam the electrical resistivity. As the temperature decreases, $\rho(T)$ curves of both sample 1 and 2 increase slightly and indicate a sudden jump below $T_{s}$ = 136~K, determined by the peak position in $d\rho(T)/dT$. At $T_{s}$, both samples indicate a hysteresis while warming and cooling the temperature, implying a first order phase transition. For clarity, the hysteresis is shown only for sample 2 in the figure. Below $T_{s}$, the $\rho(T)$ curves show a sample dependence. $\rho(T)$ of sample 1 increases continuously with a hump around 60~K and indicates a sharp drop below 4~K as a signature of AFM ordering, which is consistent with $M/H$ and $C_{p}$ measurements. The AFM phase transition temperature is determined to be $T_{N} = 4$~K from the peak position in d$\rho$/d$T$ analysis. The absolute value and temperature dependence of the resistivity for sample 1 are similar to that of CVT grown Ce$_{3}$Cd$_{2}$As$_{6}$ \cite{Piva2021, Piva2022}, where the origin of the sudden jump at 137~K in $\rho(T)$ is suggested to be associated with the CDW transition. In contrast to sample 1, $\rho(T)$ of sample 2 shows a maximum near 120 K and decreases continuously. Although sample 1 and 2 show a different temperature dependence below $T_{s}$, both samples clearly indicates the anomaly at 4~K. The inset of Fig. \ref{FIG7}(a) shows $\rho(T)$ curves of sample 2 at various magnetic fields, where $T_N$ can be suppressed below 1.8~K at $\sim$60~kOe. It should be emphasized that the electrical resistivity does not show a behavior typical of the Kondo lattice compounds. Specifically, no minimum or logarithmic upturn in $\rho(T)$ is observed.

The enhancement of $\rho(T)$ below $T_{s}$ suggests a low carrier concentration in this system. As a further probe of the carrier density, $\rho_{H}$ has been measured as a function of magnetic field and temperature. The $\rho_{H}$ curves are basically linear in field and negative above 50~K as shown in the inset of Fig.~\ref{FIG7}(b). $R_{H}$ at $H$ = 90 kOe is plotted in Fig.~\ref{FIG7}(b), where $R_{H}$ is effectively temperature independent above $T_{s} = 137$~K and indicates a sharp drop below $T_{s}$. Based on the one-band model approximation, the carrier density is estimated to be $\sim$4.87 $\times$ 10$^{22}$ per cm$^3$ at 300~K and 1.62$\times10^{19}$ per cm$^3$ at 50 K. The negative sign of $R_{H}$ indicates that transport is dominated by electron-like carriers. Figure~\ref{FIG7}(c) shows $S(T)$ of CeCd$_{0.67}$As$_{2}$, where the sample 2 of the resistivity measurement was used for the TEP measurements. $S(T)$ depends weakly on temperature above $T_{s}$ and shows a minimum near 120~K and a hump near 50~K. At low temperatures, $S(T)$ shows a sign change below $\sim$4~K, coinciding with the AFM ordering temperature. Above $T_{N}$, $S(T)$ can be described by $aT + bT^{3}$ relation, as shown in the inset. Since CeCd$_{0.67}$As$_{2}$ has magnetic Ce$^{3+}$ ions, in addition to the diffusion and phonon drag, a contribution from the CEF to the TEP should be considered. In general, many Ce- and Yb-based Kondo lattice compounds show the extrema in $S(T)$ with enhanced value due to the Kondo effect in conjunction with CEF effect. However, because $S(T)$ of CeCd$_{0.67}$As$_{2}$ shares a common temperature dependence with that of LaCd$_{0.67}$As$_{2}$ and the electrical resistivity shows no clear signature of Kondo contributions, we suggest that the broad minimum around 125~K is not associated with the Kondo effect.

\begin {figure}
\includegraphics[width=1\linewidth]{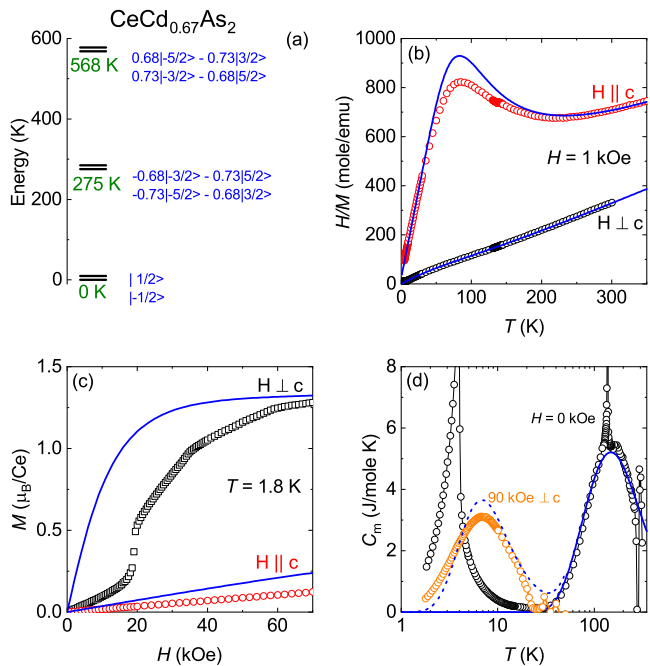}
\caption {(a) Schematic view of CEF energy levels and states. (b) Magnetic susceptibility at 1~kOe. (c) Magnetization at 1.8~K. (d) Specific heat at $H$ = 0 and 90~kOe for $H \perp c$. Open symbols represent experimental data and lines are from the CEF model (see text for details).}
\label{FIG8}
\end{figure}

For CeCd$_{0.67}$As$_2$, the strong magnetic anisotropy, broad maximum in the susceptibility for $H \parallel c$, reduced magnetization value at 70~kOe, and broad maximum in the specific heat suggest a large energy level splitting of $J = 5/2$ Hund's rule ground state of Ce-ions. In the paramagnetic state the magnetic susceptibility curves are analyzed on the basis of the CEF (point charge) model~\cite{Scheie2021}. Note that we assume the Hamiltonian below $T_{s}$ remains the same, since the magnetic susceptibility changes a little below $T_{s}$ and the in-plane magnetic anisotropy is small. The CEF Hamiltonian for the tetragonal symmetry can be expressed as ${\cal H}_{CEF} = B^{0}_{2} O^{0}_{2} + B^{0}_{4} O^{0}_{4} + B^{4}_{4} O^{4}_{4}$, where $B^{m}_{n}$ and $O^{m}_{n}$ are the CEF parameters and Stevens operators~\cite{Hutchings1964, Stevens1952}, respectively. In the presence of an applied magnetic field, the Zeeman interaction must be included: ${\cal H}_{Zeeman} = -g_{J}\mu_{B}J_{i} H_{i}$, where $g_{J}$ is the Land\'e $g$-factor, $\mu_B$ is the Bohr magneton, $H_i$ is the applied magnetic field, and $J_{i}$ is the angular momentum operator in the $i = x,y,z-$components. $M(H)$ for $H \parallel c$ and $C_{m}$ at 90~kOe cannot be reproduced by using these two terms in the Hamiltonian. Therefore, the molecular field terms are added into the total Hamiltonian: ${\cal H} = {\cal H}_{CEF} - g_{J}\mu_{B}J_{i} (H_{i} + \lambda_{i}M_{i})$, where $\lambda_{i}$ is the molecular field parameter and $M_i$ is the magnetization. The eigenvalues $E_{n}$ and the eigenfunctions $|n>$ are determined by diagonalizing the total Hamiltonian. The CEF parameters are initially obtained by fitting the magnetic susceptibility curves and determined by reproducing the measured specific heat and magnetization isotherms. From the analysis, it is found that the CEF ground state in CeCd$_{0.67}$As$_{2}$ is $|\pm 1/2>$, and the first and second excited levels are dominated by a combination of $|\pm 3/2>$ and $|\pm 5/2>$ states, as depicted schematically in Fig.~\ref{FIG8}(a).

The solid and dashed lines in Figs.~\ref{FIG8}(b)-(d) are the calculated magnetic susceptibility, magnetization, and specific heat with the following fitted parameters: $B^{0}_{2} = 19.35$~K, $B^{0}_{4} = -1.05$~K, $B^{4}_{4} = -5.46$~K, $\lambda_{\perp c} = 0.98$ mole/emu, and $\lambda_{\parallel c} = -22.82$ mole/emu. The observed anisotropy in $M(H)$ at 1.8~K is captured by the CEF model as shown in Fig.~\ref{FIG8}(b). Obviously, the calculation does not capture the MM transitions. The magnetic specific heat above 30~K is well explained by the CEF model as shown in Fig.~\ref{FIG8}(d). The energy level splitting $\Delta_{1} = 275$~K and $\Delta_{2} = 568$~K from the ground state doublet to the first and second excited doublet is consistent with the entropy change shown in Fig.~\ref{FIG6}. In particular, for $H = 90$~kOe, the low temperature broad maximum is reproduced, giving rise to a split of ground state doublet via Zeeman effect, implying well localized 4$f$ moments in CeCd$_{0.67}$As$_{2}$. It is clear from the analysis that in the paramagnetic state the magnetic property of CeCd$_{0.67}$As$_{2}$ can be well understood in the framework of CEF. In addition, the CEF analysis provides an evidence that the low temperature magnetism of CeCd$_{0.67}$As$_{2}$ can be described by effective spin $J = 1/2$. It should be noted that due to the strong CEF contribution in the magnetic susceptibility curves the obtained $\theta_{p}$ values from the high temperature Curie-Weiss fit cannot be used to estimate the size of the magnetic interaction. The unusually large molecular field term $\lambda_{\parallel c}$ implies a strong intra-layer ferromagnetic exchange couplings. In the previous study, the CEF analysis of CVT grown Ce$_{3}$Cd$_{2}$As$_{6}$ was performed by including spin exchange interaction terms (with AFM contribution for $H \perp c$ and ferromagnetic (FM) for $H \parallel c$), but only reproducing the MM transition at 34~kOe~\cite{Piva2022}. It seems that the large intra-layer FM coupling plays a crucial role in the MM transitions.

\begin {figure}
\includegraphics[width=1\linewidth]{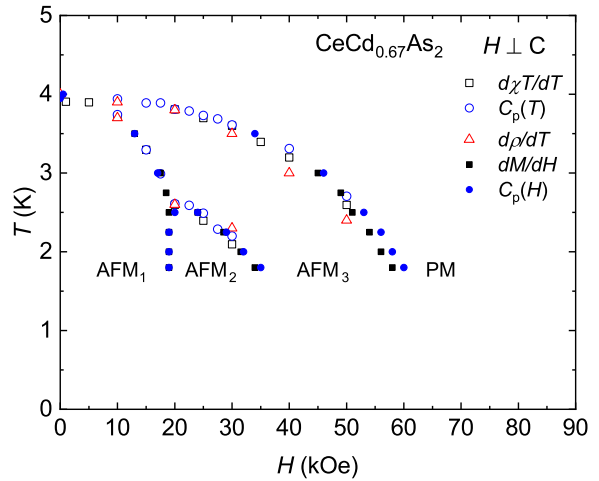}
\caption {$H-T$ phase diagram of CeCd$_{0.67}$As$_{2}$ for $H \perp c$.}
\label{FIG9}
\end{figure}

Figure~\ref{FIG9} shows the phase transition temperatures and fields determined from magnetization, resistivity, and specific heat measurements, where symbols are extracted from $d\chi T/dT$, $d\rho/dT$, $dM/dH$, and $C_{p}(T, H)$. As shown in the phase diagram the critical temperature determined from temperature sweeps track well the critical field determined from field sweeps. The phase transition can be traced by a line which is connected from $T_{N} = 4$~K at $H = 0$ to the critical field $H_{c} \sim 60$~kOe at $T = 1.8$~K for $H \perp c$. Additional phase boundaries emerge inside the AFM boundary as magnetic field increases. In spite of the abrupt changes in magnetization isotherms, no hysteresis is detected at the phase boundaries. Although further measurements below 1.8~K are required, no non-Fermi liquid behavior is observed near the critical field due to a negligible Kondo effect. The extremely low carrier density probably due to the distortion of As-square net, leaving only a few carriers in the conduction band, prevents the formation of a Kondo singlet. By doping study, it would be interesting to investigate whether the Kondo effect turns on as increasing the concentration of charge carriers. Optical tuning of wide-range carrier density is an alternative way to observe photoinduced Kondo effect as observed in the CeZn$_{3}$P${_3}$ semiconductor \cite{Kitagawa2016}. Nevertheless, RKKY interaction may be responsible for the AFM ordering, however it would have to be mediated by an extremely low charge carriers. When pressure is applied, the electrical resistivity recovers a metallic behavior and the magnetic ordering temperature increases \cite{Piva2021}. The enhancement of $T_{N}$ under pressure can be naturally explained as arising from increased carrier concentrations.

The resistivity of LaCd$_{0.67}$As$_{2}$ shows an anomaly around $T_{s} = 280$~K, which most likely has the same origin as the $T_{s} = 137$~K anomaly seen in CeCd$_{0.67}$As$_{2}$. By assuming that the high temperature transition is caused by the formation of CDW, suggested in Ref.~\cite{Piva2022}, an interesting question arises as to how the CDW and long range magnetic order interact or compete. It is necessary to perform scattering experiments to identify the origin of $T_{s}$ as well as the interaction between CDW and long range magnetic order. When electrical resistivity and Hall resistivity are compared, a relatively high mobility of charge carriers ($\sim$240 cm$^{2}$/Vs at 50~K) suggests unusual electronic state below $T_{s}$.

\section{Summary}

We have studied thermodynamic and transport properties of the single crystals of $R$Cd$_{0.67}$As$_{2}$ ($R$ = La and Ce). LaCd$_{0.67}$As$_{2}$ shows a (structural) phase transition at $T_{s}\sim280$~K below which transport properties are governed by an extremely low carrier density. CeCd$_{0.67}$As$_{2}$ exhibits a phase transition at 137~K accompanied by semiconductor-like enhancement of the electrical resistivity and indicates an AFM phase transition at 4~K. CeCd$_{0.67}$As$_{2}$ manifests a large magnetic anisotropy. The specific heat data, together with the anisotropic magnetization, was analyzed on the basis of the point charge model of crystalline electric field. In the paramagnetic phase, the observed magnetic behaviors are well explained by the CEF effects. Although the AFM phase transition temperature in CeCd$_{0.67}$As$_{2}$ can be continuously suppressed by the magnetic field to zero, the highly localized 4$f$ moments with extremely low charge carriers prevent us to observe quantum critical fluctuations shown in many Ce-based materials.

\begin{acknowledgments}
This work was supported by the Canada Research Chairs, Natural Sciences and Engineering Research Council of Canada, and Canada Foundation for Innovation program. EM was supported by the Korean Ministry of Science and ICT (No. 2021R1A2C2010925) and by BrainLink program funded by the Ministry of Science and ICT (2022H1D3A3A01077468) through the National Research Foundation of Korea.
\end{acknowledgments}

\end{document}